\documentstyle[12pt]{article}

\author{Sushkov S V\thanks{e-mail: sushkov@univex.kazan.su}\\
{\em Department of Geometry, Kazan State Pedagogical Institute,}\\
{\em Mezhlauk, 1, Kazan 420021, Tatarstan, RUSSIA}}
\title{Quantum complex scalar field in the two-dimensional spacetime
with closed timelike curves and a time-machine problem
}

\catcode`\@=11

\def\Let@{\relax\iffalse{\fi\let\\=\cr\iffalse}\fi}
\def\vspace@{\def\vspace##1{\crcr\noalign{\vskip##1\relax}}}
\def\multilimits@{\bgroup\vspace@\Let@
 \baselineskip\fontdimen10 \scriptfont\tw@
 \advance\baselineskip\fontdimen12 \scriptfont\tw@
 \lineskip\thr@@\fontdimen8 \scriptfont\thr@@
 \lineskiplimit\lineskip
 \vbox\bgroup\ialign\bgroup\hfil$\m@th\scriptstyle{##}$\hfil\crcr}
\def\Sb{_\multilimits@}
\def\endSb{\crcr\egroup\egroup\egroup}
\def\Sp{^\multilimits@}

\def\stackunder#1#2{\mathrel{\mathop{#2}\limits_{#1}}}

\def\limfunc#1{\mathop{\rm #1}}

\begin{document}

\maketitle
\begin{abstract}
It is considered the quantum complex scalar field which obeys the
authomorphic condition in the two-dimensional spacetime with closed timelike
curves and the chronology horizon. The renormalized stress-energy tensor is
obtained. It is shown that the value of the stress-energy tensor is regular
at the chronology horizon for specific authomorphic parameters. Thus the
particular example of field configuration is given for which the Hawking's
chronology protection conjecture is violated.
\end{abstract}

\section{Introduction}

The important feature of general relativity is the possible existence of
spacetimes with a nontrivial topological and causal global structure. In
particular the spacetimes can exists which have closed timelike curves
(CTC's) in a bounded region of the spacetime, such as the wormhole model
\cite{1}, the Misner universe \cite{2}, and the Gott spacetime of two moving
infinite cosmic strings \cite{3}. In such a spacetime the chronology horizon
exists. It separates chronal regions of spacetime, which are free of CTC's,
and nonchronal ones, which have CTC's. The chronology horizon is a special
type of Cauchy horizon (see for more detailes Ref.\cite{4}). The spacetime
with CTC's is multiply connected.

In the spacetime with chronology horizon the creation of time machine is
possible \cite{1} because the causality will be violated for travellers
using CTC's. To prevent the violation of causality, Hawking has proposed the
chronology protection conjecture which states that the laws of physics
prevent the formation of CTC's \cite{5}. At present it seems that the best
mechanism providing the chronology protection is the possible quantum
instability of a chronology horizon. Note that the complete solution of the
problem of chronology horizon stability is only possible in the framework of
quantum gravity. However, while the theory of quantum gravity has not
builded, one can use the semiclassical approximation to examine the
behaviour of quantum fields near the chronology horizon. The different cases
of quantum scalar fields in various spacetimes with CTC's and chronology
horizons has been successfully studied in series papers [6-13]. The general
result, which was obtained there, is that the renormalized vacuum
stress-energy tensor of scalar field is diverged at the chronology horizon.
This result is interpreted as pointing out to the quantum instability of the
chronology horizon, that can prevent the formation of CTC's.

We may ask a question: How general is the conclusion about divergent
character of the vacuum stress-energy tensor near the chronology horizon? In
order to answer to this question we must consider all kinds of matter
fields. Previously the real nontwisted scalar field had been investigated.
Now I want to consider a complex scalar field.

In this paper we will consider the quantum field theory of complex massless
scalar field in two-dimensional model of spacetime with the chronology
horizon.

In Sec.2 I give some basic ideas about a covering-space approach for
studying the quantum field theory in multiply connected spacetimes [14-18],
and show that the complex scalar field has to obey the generalized periodic
(or authomorphic) conditions in these spacetimes. The Frolov's model of
two-dimensional locally static multiply connected spacetime \cite{6} is
given in Sec.3. The vacuum stress-energy tensor for a complex scalar field
is computed in Sec.4, and it is analized in Sec.5. In Sec.6 main conclusions
are discussed. The units $c=\hbar =G=1$ are used through the paper.

\section{Field the\-ory in multiply con\-nec\-ted space\-time}

The covering-space method developed by Schulman, Dowker and Banach [14-18]
is convenient to build a quantum field theory in a multiply connected
spacetime. The main idea of this method is to regard a field theory on a
multiply connected spacetime $M$ as a field theory on the universal covering
space $\widetilde{M}$ obeying certain conditions. Let us briefly consider
some basic details of the covering-space approach.

Consider a multiply connected manifold $M$. Let $\widetilde{M}$ is an
universal (i.e., simply connected) covering space of $M$. $\Gamma $ is a
discrete group of isometry on $\widetilde{M}$ which is isomorphic to the
fundamental group $\pi _1(M)$ on $M$. $\gamma $ is elements of $\Gamma $.
The quotient $\widetilde{M}/\Gamma $ coincides with the original spacetime $%
M $.

Now let us consider the free physical field $\phi $ described by the
Lagrangian ${\cal L}[\phi (X)]{\cal .\ }$In order to quantize the field $%
\phi $ in $M$ we will use the covering-space method, i.e., we will assume
that the field $\phi $ is definded on $\widetilde{M}$ and obeyed a certain
condition. This condition is obtained from demand that the Lagrangian is
invariant under the symmetry transformations $\gamma $ of spacetime $
\widetilde{M}$:

\begin{equation}
\label{1}\gamma {\cal L}[\phi (X)]={\cal L}[\phi (\gamma X)]={\cal L}[\phi
(X)]\ .
\end{equation}
The Lagrangian of free fields is quadratic in $\phi $ so the next field
transformation follows from (1)
\begin{equation}
\label{2}\phi (\gamma X)=a(\gamma )\phi (X)
\end{equation}
where $a^2(\gamma )=1$ and $a(\gamma _1\gamma _2)=a(\gamma _1)a(\gamma _2)$
(the group property). (For the sake of simplicity we assume that $a(\gamma )$
does not depend on a point $X$, what means the global symmetry.) The fields $%
\phi $ on $\widetilde{M}$ obeing Eq.(2) are known as authomorphic fields and
the condition (2) is called the authomorphic one.

The Lagrangian of conformally coupled complex massless scalar field has the
form

\begin{equation}
\label{3}{\cal L}=-g^{\mu \nu }\phi _{,\mu }\phi _{,\nu }^{*}-\xi R\phi \phi
^{*}\ ,
\end{equation}
where $R$ is a scalar curvature and the parameter $\xi $ is equal to $\frac
16$ or $0$ for four- or two-dimensional cases correspondingly. The
authomorphic condition (\ref{2}) for complex fields reads
\begin{equation}
\label{4}\phi (\gamma X)=e^{2\pi i\alpha }\phi (X)\,\,,
\end{equation}
where $\alpha $ is an arbitrary real parameter, $0\leq \alpha \leq 1/2$ (we
will call it as an authomorphic parameter). There are two particular special
cases $\alpha =0$ and $\alpha =1/2$ for which the field transformation rule (%
\ref{4}) takes forms $\phi (\gamma X)=\phi (X)$ (nontwisted field) and $\phi
(\gamma X)=-\phi (X)$ (twisted field).

So, below we will study the theory of authomorphic field $\phi $ on the
universal covering space $\widetilde{M}$, which obeys the authomorphic
condition (\ref{4}), instead of the complex field theory on $M$.

\section{Two-dimen\-sional model of spacetime with chronology horizon}

In order to study the behaviour of complex field near the chronology horizon
we will take the two-dimensional model of spacetime which was proposed by
Frolov for investigation of the ordinary (i.e., real nontwisted) massless
scalar field. Let us consider this model. (For more detailes, see Ref.\cite
{6}.) It is the two-dimensional locally static spacetime with the metric
\begin{equation}
\label{5}ds^2=-e^{-2Wl}dt^2+dl^2\ \,,
\end{equation}
where $W=const$ is a parameter, $t\in (-\infty ,\infty )$, the proper
distance coordinate $l$ changes from $l=0$ to $l=L$, and the boundary points
$0$ and $L$ being considering identical. The regularity of the spacetime $M$
requires that the metric (\ref{5}) is identical at both boundaries: $\gamma
^{-},\,l=0$ and $\gamma ^{+},\,l=L$. It is guaranteed if the time parameters
of point $(t,0)$ of $\gamma ^{-}$ and $(t^{\prime },L)$ of $\gamma ^{+}$,
which are to be identified, are related as follows:
\begin{equation}
\label{6}t^{\prime }=At\ ,\ A\equiv e^{WL}\ .
\end{equation}
Therefore, the next identification rule: $(t,0)\longleftrightarrow (At,L)$
is assumed.

In $M$ the local (defined in any simply connected region $U\subset M$)
uniquely defined (up to normalization) nonvanishing timelike Killing vector
field $\xi ^\mu (1,0)$ exists. But it is not defined globally. Really, the
norm of Killing vector $\left| \xi ^2\right| ^{1/2}=e^{-Wl}$ cannot be fixed
in a whole of spacetime $M$. It changes on the quantity $A=e^{WL}$ in one
pass along a closed spatial path, where the coordinate $l$ changes from $0$
to $L$. The definition of gravitational potential $\varphi $ in the
spacetime $M$ reads
\begin{equation}
\label{7}e^\varphi =\left| \xi ^2\right| ^{1/2}=e^{-Wl},
\end{equation}
and it is clear that the potential $\varphi $ cannot be also defined
globally in $M$. Such gravitational field is called nonpotential. The degree
of nonpotentiality is characterized by the value $A$.

For looking into the causal structure of $M$ let us take null coordinates
\begin{equation}
\label{8}u=W^{-1}e^{Wl}-t\,,\,\,\,v=W^{-1}e^{Wl}+t\,.
\end{equation}
The equations $u=const$ and $v=const$ determinate null geodesics in
spacetime $M$. The lines $u=0$ and $v=0$ are closed null geodesics. In the
chronal region $R_{+}:uv>0$ there are no CTC's, while CTC's are possible in
the nonchronal regions lying beyond $R_{+}$. Hence lines $u=0$ and $v=0$
form chronology horizons $H_{+}$ and $H_{-}$ correspondingly.

\section{Vacuum stress-energy tensor}

Let us begin now the calculation of renormalized stress-energy tensor for
the complex scalar field. The universal covering space $\widetilde{M}$ for $%
M $ is a spacetime with the metric
\begin{equation}
\label{9}d\tilde s^2=-e^{-2Wl}dt^2+dl^2
\end{equation}
where $t\in (-\infty ,\infty )\ ,l\in (-\infty ,\infty )$. By using the
dimensionless coordinates
\begin{equation}
\label{10}\eta =Wt;\ \xi =\exp (Wl),
\end{equation}
this metric can be rewritten in the form
\begin{equation}
\label{11}d\tilde s^2=(W\xi )^{-2}\,(-d\eta ^2+d\xi ^2)\ .
\end{equation}
The action of operator $\gamma _n\equiv (\gamma )^n$ (where $\gamma $ is
generators of the fundamental group $\Gamma $) on $\tilde M$ is described by
the relations
\begin{equation}
\label{12}\gamma _n\eta =A^n\eta ,\ \gamma _n\xi =A^n\xi \ .
\end{equation}
The strip $\xi \in (1,A),\ \eta \in (-\infty ,\infty )$ is a fundamental
domain.

Consider in $\widetilde{M}$ a complex massless scalar field $\phi $ with the
Lagrangian (\ref{3}) obeying the wave equation

\begin{equation}
\label{13}\Box \phi =0\ .
\end{equation}
(For the complex conjugated field $\phi ^{*}$ the wave equation is the same,
so, as usually, we will consider the case of $\phi $ only.) In null
coordinates
\begin{equation}
\label{14}\zeta _{-}\equiv Wu=\xi -\eta ;\ \zeta _{+}\equiv Wv=\xi +\eta
\end{equation}
the field equation (\ref{13}) reads
\begin{equation}
\label{15}\partial _{+-}\phi =\partial _{-+}\phi =0\ ,
\end{equation}
where $\partial _{+-}=\partial _{+}\partial _{-}$ and $\partial _{\pm
}=\partial /\partial \zeta _{\pm }$. A general positive-frequency solution
of this equation can be written as follows
\begin{equation}
\label{16}u(\eta ,\xi )=\int_0^\infty \frac{d\omega }\omega a(\omega )\left[
e^{i\omega \zeta _{-}}-e^{-i\omega \zeta _{+}}\right] \ .
\end{equation}
This solution provides a fulfillment of the boundary condition

\begin{equation}
\label{17}u_{\xi =0}=0\,.
\end{equation}
(It is easy to show that regular solutons of massive scalar field equation
vanish at $\xi =0$ for arbitrary small value of mass.)

The positive-frequency solutions

\begin{equation}
\label{18}U_\omega =(4\pi \omega )^{-1/2}\left[ e^{i\omega \zeta
_{-}}-e^{-i\omega \zeta _{+}}\right]
\end{equation}
obey the normalizaton conditions

\begin{equation}
\label{19}\left\langle U_\omega U_{\omega ^{\prime }}\right\rangle
=-i\int\Sb \eta =const \\ \xi >0\endSb \left( U_\omega \stackrel{.}{
\overline{U}}_{\omega ^{\prime }}-\stackrel{.}{U}_\omega \overline{U}%
_{\omega ^{\prime }}\right) d\xi =\delta (\omega -\omega ^{\prime }),
\end{equation}
where the dot means $d/d\eta $, and form a basis in $H_{\widetilde{M}}$.

The Hadamard function $\widetilde{G}^1$ in $\widetilde{M}$ is
\begin{equation}
\label{20}\widetilde{G}^1(X,X^{\prime })=\stackunder{\omega }{\sum }\left(
U_\omega (X)\overline{U}_\omega (X^{\prime })+U_\omega (X^{\prime })
\overline{U}_\omega (X)\right) =
\end{equation}
\begin{equation}
\label{21}=-\frac 1{4\pi }\ln \left( \frac{(\zeta _{-}-\zeta _{-}^{\prime
})^2(\zeta _{+}-\zeta _{+}^{\prime })^2}{(\zeta _{+}+\zeta _{-}^{\prime
})^2(\zeta _{-}+\zeta _{+}^{\prime })^2}\right) .
\end{equation}

Let us demand now that the positive-frequency solution (\ref{16}) obeys the
authomorphic condition (\ref{4}). Then, the relation
\begin{equation}
\label{22}\gamma u(\eta ,\xi )=u(\gamma \eta ,\gamma \xi )=u(A\eta ,A\xi
)=e^{2\pi i\alpha }u(\eta ,\xi )
\end{equation}
must be fulfilled. It is fulfilled provided
\begin{equation}
\label{23}a(\omega )=e^{2\pi i\alpha }a(A\omega )\,.
\end{equation}

The general solution of this functional equation is
\begin{equation}
\label{24}a(\omega )=\stackunder{n}{\sum }c_n\omega ^{-2\pi i\beta (n+\alpha
)},
\end{equation}
where $\beta \equiv (\ln A)^{-1}=(WL)^{-1}$, $c_n$ are constants, and the
summation is taken over all integer numbers $n$. By using the formula \cite
{19}%
$$
\int_0^\infty \frac{d\omega }\omega \omega ^{2\pi i\beta (n+\alpha )}e^{\pm
i\omega \zeta }=
$$
\begin{equation}
\label{25}=\pm ie^{\pm i\pi /2}e^{\mp \pi ^2\beta (n+\alpha )}\Gamma (2\pi
i\beta (n+\alpha ))(\zeta \pm i0)^{-2\pi i\beta (n+\alpha )}
\end{equation}
one can show from (\ref{16}) that the positive-frequency authomorphic
solution allows the representation
\begin{equation}
\label{26}u=\stackunder{n}{\sum }\widetilde{c}_nu_n\,,
\end{equation}
where
\begin{equation}
\label{27}u_n=b_n\left[ e^{\pi ^2\beta (n+\alpha )}(\zeta _{+}-i0)^{-2\pi
i\beta (n+\alpha )}-e^{-\pi ^2\beta (n+\alpha )}(\zeta _{-}+i0)^{-2\pi
i\beta (n+\alpha )}\right] .
\end{equation}
The scalar product for authomorphic solutions (\ref{27}), defind in $M$,
does not depend on the particular choice of the Cauchy surface in the
fundamental domain and we may choose the section $\eta =0$ as such a surface
and write
\begin{equation}
\label{28}\left\langle u^1,u^2\right\rangle =-i\int_1^Ad\xi \left( u^1%
\stackrel{.}{\bar u}^2-\stackrel{.}{u}^1\bar u^2\right) _{\eta =0}\,.
\end{equation}
The solutions (\ref{27}) form an orthonormal basis in $H_M$:
\begin{equation}
\label{29}\left\langle u_n,u_{n^{\prime }}\right\rangle =\delta _{nn^{\prime
}}
\end{equation}
provided
\begin{equation}
\label{30}b_n=\left[ 8\pi (n+\alpha )\sinh (\chi (n+\alpha ))\right]
^{-1/2}\,,
\end{equation}
where $\chi \equiv 2\pi ^2\beta $. The Hadamard function $G^1$ defined in $M$
is
\begin{equation}
\label{31}
\begin{array}{c}
G^1(X,X^{\prime })=
\stackunder{n}{\sum }\left( u_n(X)\bar u_n(X^{\prime })+u_n(x^{\prime })\bar
u_n(X)\right) = \\ =
\stackunder{n}{\sum }b_n^2\left\{ e^{\chi (n+\alpha )}\left( \bar \zeta
_{+}^{\prime }/\zeta _{+}\right) ^{i\mu (n+\alpha )}+e^{-\chi (n+\alpha
)}\left( \bar \zeta _{-}^{\prime }/\zeta _{-}\right) ^{i\mu (n+\alpha
)}-\right. \\ \left. -\left( \bar \zeta _{+}^{\prime }/\zeta _{-}\right)
^{i\mu (n+\alpha )}-\left( \bar \zeta _{-}^{\prime }/\zeta _{+}\right)
^{i\mu (n+\alpha )}\right\} +(c.c.)\,,
\end{array}
\end{equation}
where $\mu =2\pi \beta $, $\zeta _{\pm }\equiv \zeta _{\pm }\mp i0$, $\bar
\zeta _{\pm }=\bar \zeta _{\pm }\pm i0$. In the region $R_{+}$ where $\zeta
_{\pm }>0$ and $\zeta _{\pm }^{\prime }>0$, the values $\zeta _{\pm }$, $%
\zeta _{\pm }^{\prime }$ and their complex conjugate coincide. The Hadamard
function for this case can be rewritten as%
$$
G^1(X,X^{\prime })=\frac 1{4\pi }\stackunder{\epsilon =+,-}{\stackrel{\infty
}{\stackunder{n=-\infty }{\sum }}}\frac{e^{\chi (n+\alpha )}\cos (n+\alpha
)y_\epsilon }{(n+\alpha )\sinh \chi (n+\alpha )}-
$$
\begin{equation}
\label{32}-\frac 1{4\pi }\stackrel{\infty }{\stackunder{n=-\infty }{\sum }}
\frac{\cos \left[ \mu (n+\alpha )\ln (\zeta _{+}^{\prime }/\zeta
_{-})\right] }{(n+\alpha )\sinh \chi (n+\alpha )}-\frac 1{4\pi }\stackrel{%
\infty }{\stackunder{n=-\infty }{\sum }}\frac{\cos \left[ \mu (n+\alpha )\ln
(\zeta _{-}^{\prime }/\zeta _{+})\right] }{(n+\alpha )\sinh \chi (n+\alpha )}%
\,,
\end{equation}
where $y_{\pm }=\mu \ln (\zeta _{\pm }^{\prime }/\zeta _{\pm })$. Consider
separately the first term in the expression (\ref{32}). It can be given as
follows%
$$
\frac 1{4\pi }\stackunder{\epsilon =+,-}{\stackrel{\infty }{\stackunder{%
n=-\infty }{\sum }}}\frac{e^{\chi (n+\alpha )}\cos (n+\alpha )y_\epsilon }{%
(n+\alpha )\sinh \chi (n+\alpha )}=\frac 1{2\pi }\stackunder{\epsilon =+,-}{%
\stackrel{\infty }{\stackunder{n=0}{\sum }}}\frac{\cos (n+\alpha )y_\epsilon
}{(n+\alpha )}+
$$
\begin{equation}
\label{33}+\frac 1{2\pi }\stackunder{\epsilon =+,-}{\stackrel{\infty }{%
\stackunder{n=0}{\sum }}}\frac{\cos (n+\alpha )y_\epsilon }{(n+\alpha
)\left[ e^{2\chi (n+\alpha )}-1\right] }+\frac 1{2\pi }\stackunder{\epsilon
=+,-}{\stackrel{\infty }{\stackunder{n=1}{\sum }}}\frac{\cos (n-\alpha
)y_\epsilon }{(n-\alpha )\left[ e^{2\chi (n-\alpha )}-1\right] }\,.
\end{equation}
Using the formula (5.3.2) from \cite{20}%
$$
\stackrel{\infty }{\stackunder{n=0}{\sum }}\frac 1{n+\alpha }\left\{
\begin{array}{c}
\sin yn \\
\cos yn
\end{array}
\right\} =\beta (\alpha )\left\{
\begin{array}{c}
\sin (\pi -y)\alpha \\
\cos (\pi -y)\alpha
\end{array}
\right\} +
$$
\begin{equation}
\label{34}+\frac 12\stackrel{\pi }{\stackunder{y}{\int }}\left\{
\begin{array}{c}
\sin [-y\alpha +(\alpha -1/2)t] \\
\cos [-y\alpha +(\alpha -1/2)t]
\end{array}
\right\} \csc \frac t2\,dt\,,
\end{equation}
where $\beta (\alpha )=\frac 12\left[ \psi \left( \frac{\alpha +1}2\right)
-\psi \left( \frac \alpha 2\right) \right] $ ; \thinspace \thinspace $\psi
(z)=\Gamma ^{\prime }(z)/\Gamma (z)$ is a psi-function, one can rewrite the
first series in (\ref{33}) as follows
\begin{equation}
\label{35}\frac 1{2\pi }\stackunder{\epsilon =+,-}{\stackrel{\infty }{%
\stackunder{n=0}{\sum }}}\frac{\cos (n+\alpha )y_\epsilon }{(n+\alpha )}=
\frac{\beta (\alpha )}\pi \cos \pi \alpha +\frac 1{4\pi }\stackunder{%
\epsilon =+,-}{\sum }\int_{y_\epsilon }^\pi \frac{\cos (\alpha -1/2)t}{\sin
(t/2)}dt\,.
\end{equation}
Finally, the expression (\ref{32}) for the Hadamard function takes the form%
$$
G^1(X,X^{\prime })=\frac{\beta (\alpha )}\pi \cos \pi \alpha +\frac 1{4\pi }%
\stackunder{\epsilon =+,-}{\sum }\int_{y_\epsilon }^\pi \frac{\cos (\alpha
-1/2)t}{\sin (t/2)}dt+
$$
$$
+\frac 1{2\pi }\stackunder{\epsilon =+,-}{\stackrel{\infty }{\stackunder{n=0%
}{\sum }}}\frac{\cos (n+\alpha )y_\epsilon }{(n+\alpha )\left[ e^{2\chi
(n+\alpha )}-1\right] }+\frac 1{2\pi }\stackunder{\epsilon =+,-}{\stackrel{%
\infty }{\stackunder{n=1}{\sum }}}\frac{\cos (n-\alpha )y_\epsilon }{%
(n-\alpha )\left[ e^{2\chi (n-\alpha )}-1\right] }-
$$
\begin{equation}
\label{36}-\frac 1{4\pi }\stackrel{\infty }{\stackunder{n=-\infty }{\sum }}
\frac{\cos \left[ \mu (n+\alpha )\ln (\zeta _{+}^{\prime }/\zeta
_{-})\right] }{(n+\alpha )\sinh \chi (n+\alpha )}-\frac 1{4\pi }\stackrel{%
\infty }{\stackunder{n=-\infty }{\sum }}\frac{\cos \left[ \mu (n+\alpha )\ln
(\zeta _{-}^{\prime }/\zeta _{+})\right] }{(n+\alpha )\sinh \chi (n+\alpha )}%
\,.
\end{equation}

In order to calculate the renormalized stress-energy tensor one can use the
standard point-splitting method [21,22]. But in our case due to the high
symmetry of the universal covering space it is possible to simplify the
calculations. The space $M$ is locally isometric to $\widetilde{M}$. Thus
the terms which are to be subtracted in order to renormalize the vacuum
expectation value of the stress-energy tensor are the same in the both
spaces and we can write
\begin{equation}
\label{37}\left\langle T_{\mu \nu }\right\rangle _M^{ren}=T_{\mu \nu
}^1+\left\langle T_{\mu \nu }\right\rangle _{\widetilde{M}}^{ren}\,,
\end{equation}
where $\left\langle T_{\mu \nu }\right\rangle _M^{ren}$ and $\left\langle
T_{\mu \nu }\right\rangle _{\widetilde{M}}^{ren}$ are the renormalized
stress-energy tensors in $M$ and $\widetilde{M}$ correspondingly. For the
scalar field the point-splitting method gives (see Ref.\cite{21})
\begin{equation}
\label{38}\left\langle T_{\mu \nu }\right\rangle ^{ren}=\stackunder{%
X\rightarrow X^{\prime }}{\lim }{\cal D}_{\mu \nu }G^{1,ren}\,,
\end{equation}
where%
$$
{\cal D}_{\mu \nu }=\frac 14(\nabla _\mu \nabla _{\nu ^{\prime }}+\nabla
_{\mu ^{\prime }}\nabla _\nu -g_{\mu \nu }g^{\rho \sigma ^{\prime }}\nabla
_\rho \nabla _{\sigma ^{\prime }})\,.
$$
Thus we have the following expression for $T_{\mu \nu }^1$%
\begin{equation}
\label{39}T_{\mu \nu }^1=\stackunder{X\rightarrow X^{\prime }}{\lim }{\cal D}%
_{\mu \nu }[G^1(X,X^{\prime })-\widetilde{G}^1(X,X^{\prime })]\,,
\end{equation}
The expression in the square brackets on the right-hand side of (\ref{39})
is a regular function. It is easy to see that $T_{+\,-}^1=T_{-\,+}^1=0$,
where we use the notations $T_{\pm \,\mp }^1=T_{\zeta _{\pm }\zeta _{\mp
}}^1 $. While the nonvanishing components $T_{+\,+}^1$ and $T_{-\,-}^1$ are
\begin{equation}
\label{40}T_{\pm \,\pm }^1=\stackunder{X\rightarrow X^{\prime }}{\lim }%
\partial _{\pm }\partial _{\pm }^{\prime }[G^1(X,X^{\prime })-\widetilde{G}%
^1(X,X^{\prime })]\,.
\end{equation}
Differentiating the Hadamard function (\ref{36}) we obtain%
$$
\partial _{\pm }\partial _{\pm }^{\prime }G^1(X,X^{\prime })=\frac{\mu ^2}{%
\zeta _{\pm }\zeta _{\pm }^{\prime }}\left\{ -\frac 1{8\pi }\frac{\cos
(y_{\pm }/2)\cos (\alpha -1/2)y_{\pm }}{\sin {}^2(y_{\pm }/2)}-\right. \,
$$
$$
-\frac{\alpha -1/2}{4\pi }\frac{\sin (\alpha -1/2)y_{\pm }}{\sin (y_{\pm }/2)%
}+\frac 1{2\pi }\stackrel{\infty }{\stackunder{\epsilon =+,-}{\stackunder{n=0%
}{\sum }}}\frac{(n+\alpha )\cos (n+\alpha )y_\epsilon }{e^{2\chi (n+\alpha
)}-1}+
$$
\begin{equation}
\label{41}\left. +\frac 1{2\pi }\stackrel{\infty }{\stackunder{\epsilon =+,-%
}{\stackunder{n=1}{\sum }}}\frac{(n-\alpha )\cos (n-\alpha )y_\epsilon }{%
e^{2\chi (n-\alpha )}-1}\right\} \,.
\end{equation}
By using this relation and the expression (\ref{21}) for the Hadamard
function $\widetilde{G}^1(X,X^{\prime })$ in $\widetilde{M}$ one can obtain
for $T_{\mu \nu }^1$ in the limit of coinciding points $X\rightarrow
X^{\prime }$:
\begin{equation}
\label{42}T_{\pm \,\pm }^1=\frac 1{\zeta _{\pm }^2}F_\alpha (\beta )\,,
\end{equation}
where a function $F_\alpha (\beta )$ is defined as follows%
$$
F_\alpha (\beta )=-\frac 1{24\pi }+\pi \beta ^2\left[ \frac 1{12}-\left(
\alpha -\frac 12\right) ^2\right] +
$$
\begin{equation}
\label{43}+2\pi \beta ^2\stackrel{\infty }{\stackunder{n=0}{\sum }}\frac{%
n+\alpha }{e^{2\chi (n+\alpha )}-1}+2\pi \beta ^2\stackrel{\infty }{%
\stackunder{n=1}{\sum }}\frac{n-\alpha }{e^{2\chi (n-\alpha )}-1}\,.
\end{equation}

The renormalized stress-energy tensor $\left\langle T_{\mu \nu
}\right\rangle _{\widetilde{M}}^{ren}=-g_{\mu \nu }/24\pi $, where $g_{\mu
\nu }$ is the metric in $\widetilde{M}$, was obtained in Ref.\cite{6} for a
real scalar field. In the case of complex field this result must be
multiplied by two, so that
\begin{equation}
\label{43.1}\left\langle T_{\mu \nu }\right\rangle _{\widetilde{M}%
}^{ren}=-g_{\mu \nu }/12\pi \,.
\end{equation}
Combining expressions (\ref{42}) and (\ref{43.1}) we can write finally for
the renormalized stress-energy tensor of authomorphic complex scalar field:
\begin{equation}
\label{44}\left\langle T_{\mu \nu }\right\rangle _M^{ren}=\left( \frac
1{\zeta _{+}^2}\delta _\mu ^{+}\delta _\nu ^{+}+\frac 1{\zeta _{-}^2}\delta
_\mu ^{-}\delta _\nu ^{-}\right) F_\alpha (\beta )-\frac 1{12\pi }g_{\mu \nu
}\,.
\end{equation}

Before investigation of the obtained result we consider the case when the
nonpotential gravitational field is weak, i.e., $A=e^{WL}\rightarrow 1$.
Then we have $\delta \equiv A-1<<1$, $W\simeq \delta /L$ and $\beta =(\ln
A)^{-1}\simeq \delta ^{-1}>>1$. In the $\delta \rightarrow 0$ limit, the
terms of (\ref{43}) that contain series are of the order $\delta ^{-2}\exp
(-4\pi ^2/\delta )$, and it can be neglected. Thus we have $F_\alpha (\beta
)\simeq \pi \delta ^{-2}[1/12-(\alpha -1/2)^2]$. For fixed values of $l$ and
$t$, one has $u\simeq v\simeq L/\delta $, and hence up to the terms which
vanish in the $\delta \rightarrow 0$ limit one has
\begin{eqnarray}
\label{45}
\left\langle T_{uu}\right\rangle _M^{ren} & = & \left\langle
T_{vv}\right\rangle _M^{ren} =\frac \pi {L^2}\left[ \frac 1{12}-\left(
\alpha -\frac 12\right) ^2\right] \,; \nonumber \\
\left\langle T_{uv}\right\rangle _M^{ren} & = & \left\langle
T_{vu}\right\rangle _M^{ren} =0 \,.
\end{eqnarray}
These expressions correctly reproduce the value of the renormalized
stress-energy tensor for the authomorphic complex conformal massless scalar
field in a two-dimensional cylindrical spacetime (see, e.g., Ref.\cite{23}).

In particular cases when $\alpha =0$ (the ordinary field) and $\alpha =1/2$
(the twisted field) we have $\left\langle T_{uu}\right\rangle
_M^{ren}=\left\langle T_{vv}\right\rangle _M^{ren}=-\pi /6L^2$ and $+\pi
/12L^2$. As one can see the value of stress-energy tensor changes himself
sign depending on values of the authomorhic parameter. Note also that the
special value
\begin{equation}
\label{46}\alpha _0=\frac 12-\frac 1{\sqrt{12}}
\end{equation}
exists for which the stress-energy tensor (\ref{45}) is equal to zero for
any $L$.

\section{The behaviour of stress-energy tensor near the chronology horizon}

Consider now the general case of nonpotential gravitational field with an
arbitrary value of $\beta $. As it was mentioned before the equations $u=0$
and $v=0$ describe closed null geodesics which form the chronology horizons $%
H_{+},\,H_{-}$ . The leading term of $\left\langle T_{\mu \nu }\right\rangle
_M^{ren}$ near the chronology horizon $H_{+}:\,\zeta _{-}=0$ is
\begin{equation}
\label{47}\left\langle T_{uu}\right\rangle _M^{ren}=\frac{F_\alpha (\beta )}{%
u^2}k_\mu k_\nu \,,\,\,\,k_\mu =\nabla _\mu \,u\,
\end{equation}
if only the function $F_\alpha (\beta )$ is {\em not }became equal to zero
(this case we will consider especially). Then, the expression (\ref{47})
shows that near $H_{+}$ there exists an infinitely growing flux of energy
density. The sign of energy density is determined by the sign of the
function $F_\alpha (\beta )$. So let us investigate it in details.

At first we take the case $\alpha =0$ corresponding to the nontwisted scalar
field. The next expression for $F_\alpha (\beta )$ follows from (\ref{43}):
\begin{equation}
\label{48}F_{\alpha =0}(\beta )=-\frac 1{24\pi }+\frac \beta {2\pi }-\frac{%
\pi \beta ^2}6+4\pi \beta ^2\stackrel{\infty }{\stackunder{n=0}{\sum }}%
\,\frac n{e^{2\chi n}-1}\,.
\end{equation}
It coincides with the formula for $F_{\alpha =0}(\beta )$ obtained in \cite
{6}. It was also shown in \cite{6} that the function $F_{\alpha =0}(\beta )$
allows another representation:
\begin{equation}
\label{49}F_{\alpha =0}(\beta )=-\frac 1{4\pi }\stackrel{\infty }{%
\stackunder{n=0}{\sum }}\,\frac 1{\sinh {}^2(n/2\beta )}\,,
\end{equation}
from which one can clearly see that $F_{\alpha =0}(\beta )$ is a negative
defined function for all values of $\beta $. It means that the renormalized
stress-energy tensor of the scalar field with $\alpha =0$ is divergent near
the chronology horizon for any value of nonpotentiality of the gravitational
field. And then, the infinitely growing flux of negative energy density
appears which is propagating in the direction of the gravitational potential
decrease.

Consider now the case $\alpha =1/2$ corresponding to the twisted scalar
field. We have from (\ref{43}):
\begin{equation}
\label{50}F_{\alpha =1/2}(\beta )=-\frac 1{24\pi }+\frac{\pi \beta ^2}{12}%
+4\pi \beta ^2\stackrel{\infty }{\stackunder{n=0}{\sum }}\,\frac{n+1/2}{%
e^{2\chi (n+1/2)}-1}\,.
\end{equation}
As it is shown in the appendix A the function $F_{\alpha =1/2}(\beta )$ can
be rewritten in another equivalent form
\begin{equation}
\label{51}F_{\alpha =1/2}(\beta )=-\frac 1{24\pi }+\frac 1{3\pi }K^2(m)\beta
^2(m+1)\,,
\end{equation}
where $m$ is a parameter, $0\leq m\leq 1$; $K(m)$ is an elliptic integral
\cite{24}: $K(m)=\int_0^{\pi /2}d\theta (1-m\sin {}^2\theta )^{-1/2}\,.$ The
parameter $m$ is connected with the nonpotentiality parameter $\beta $ as
follows (see appendix A): $2\pi \beta =K(1-m)/K(m)$. It is also shown in
appendix A that the function $F_{\alpha =1/2}(\beta )$ defined by the
relation (\ref{51}) is positive for all values of $\beta $. It means that
the renormalized stress-energy tensor of twisted scalar field is divergent
near the chronology horizon for any values of nonpotentiality of a
gravitational field. And then, the infinitly growing flux of positive energy
density appears which is propagating in the direction of the gravitational
potential increase.

Thus we have obtained that the function $F_\alpha (\beta )$ is negative
defined for the ordinary scalar field and positive defined for the twisted
one. As the consequence of this fact the infinitly growing flux of negative
or positive energy density exists near the chronology horizon. This result
can be interpreted as an evidence of quantum nonstability of the chronology
horizon.

In the case of complex authomorphic scalar field the parameter $\alpha $ can
take values within the interval $[0,1/2]$. The function $F_\alpha (\beta )$
is negative or positive defined for the extreme values $\alpha =0$ and $%
\alpha =1/2$ correspondingly. It is clear that the value $\alpha $ must
exist when $F_\alpha (\beta )$ is equal to zero (at least for some $\beta $%
). In order to analize the behaviour of function $F_\alpha (\beta )$ we will
build it graph for different values of $\alpha $. The form of $F_\alpha
(\beta )$ is defined by the expression (\ref{43}). The series in this
expression is fast convergent for large values of $\beta $ and slowly
convergent for small ones. In the appendix B the other representation of $%
F_\alpha (\beta )$ is obtained which is convenient for the numerical
analysis in the case of small $\beta $:
\begin{equation}
\label{52}F_\alpha (\beta )=-\frac 1{4\pi }\stackrel{\infty }{\stackunder{n=1%
}{\sum }}\frac{\cos 2\pi n\alpha }{\sinh {}^2(n/2\beta )}\,.
\end{equation}
The graph of function $F_\alpha (\beta )$ defined by the formulae (\ref{43})
and (\ref{52}), is given in Fig.1.

\begin{figure}
\vspace{5cm}
\caption
{The graph of $F_\alpha(\beta)$. The curve {\bf a} corresponds to the value
 $0\leq \alpha < \alpha_0$; the curve {\bf b} is obtained if $\alpha =
\alpha_0$;
 the curves {\bf c}, {\bf d} are obtained if $\alpha_0 <\alpha < 1/2$;
 and the curve {\bf e} corresponds to the value $\alpha = 1/2$.}
\end{figure}

 From Fig.1 one can see the next characteristic features of the behaviour of
the function $F_\alpha (\beta )$. If $0\leq \alpha <\alpha _0,\,\,\alpha
_0=\frac 12-\frac 1{\sqrt{12}}$ then $F_\alpha (\beta )$ is less than zero
for all values of $\beta $. In the special case $\alpha =\alpha _0$ the
graph of $F_\alpha (\beta )$ is asymptotically approached to the straight
line $-\frac 1{24\pi }$. If $\alpha =\alpha _{*}$, where $\alpha _0<\alpha
_{*}<1/2$, then $F_\alpha (\beta )$ is a function with alternating signs; it
is less than zero if $0<\beta <\beta _{*}$ and greater than zero if $\beta
>\beta _{*}$. At the point $\beta =\beta _{*}$ the function $F_\alpha (\beta
)$ is equal to zero; the value $\beta _{*}$ is defined from the equation $%
F_{\alpha _{*}}(\beta _{*})=0$. And, at last, $F_\alpha (\beta )$ is greater
than zero everywhere for $\alpha =1/2$. (Note that at the point $\beta =0$
the function $F_\alpha (\beta )$ is equal to zero for all values of
parameter $\alpha $. The value $\beta =0$ corresponds to the infinite
nonpotentiality of gravitational field, $W\rightarrow \infty $.)

Now we can do the complete characterization of stress-energy tensor (\ref{44}%
) near the chronology horizon. There are two qualitatively different cases.
{\bf (i)} The first case, when the function $F_\alpha (\beta )$ is not equal
to zero, then the leading term in expression (\ref{44}) near the chronology
horizon takes the form (\ref{47}), and at the chronology horizon the
stress-energy tensor is divergent. For various values $\alpha $ and $\beta $
the function $F_\alpha (\beta )$ can be negative or positive. In consequence
of this fact the infinitly growing flux of negative or positive energy
density exists near the horizon. {\bf (ii)} And second case, when $F_\alpha
(\beta )$ is equal to zero. It is possible if the authomorphic parameter $%
\alpha =\alpha _{*}$ takes its values within the interval $(\alpha _0,1/2)$,
then such parameter $\beta =\beta _{*}$ exists that $F_{\alpha _{*}}(\beta
_{*})=0$. In this case the expression (\ref{44}) for $\left\langle T_{\mu
\nu }\right\rangle _M^{ren}$ takes the form
\begin{equation}
\label{53}\left\langle T_{\mu \nu }\right\rangle _M^{ren}=-g_{\mu \nu
}/12\pi ,
\end{equation}
and one can see that the stress-energy tensor is regular at the chronology
horizons $H_{+}$ and $H_{-}$.

\section{Summary and concluding remarks}

In this paper we have considered the complex massless scalar field in
two-dimensional model of spacetime $M$ with closed timelike curves. We have
obtained the expression for renormalized vacuum stress-energy tensor of
scalar field (see Eq.(\ref{44})) which depends on the authomorphic parameter
$\alpha $, defining the transformation law of scalar field under the
symmetry transformations of spacetime $M$, and the parameter $\beta $,
characterizing the degree of nonpotentiality of the gravitational field. The
analysis of behaviour of $\left\langle T_{\mu \nu }\right\rangle _M^{ren}$
near the chronology horizon shows that there are two qualitatively different
cases. In first case (in particular, when $\alpha =0$ (nontwisted field) and
$\alpha =1/2$ (twisted field)) the value of stress-energy tensor is diverged
near the chronology horizon and the infinitly growing flux of negative or
positive energy density exists there. This fact can be interpreted as
pointing out to the quantum instability of the chronology horizon. Note that
the obtained result extends conclusions of works [6-13] for the case of
complex field.

The essentially new important result, which is obtained in this paper, is
the fact that the stress-energy tensor of authomorphic scalar field is
finite at the chronology horizon for some values of parameters $\alpha $ and
$\beta $. It means that the formation of chronology horizon is possible for
this field configuration. Hence, the Hawking's chronology protection
conjecture does not work for the case of authomorphic fields.

A question remains: How can we determine and fix the values of parameters $%
\alpha $ and $\beta $? I think that the answer is only possible in the
framework of selfconsistent investigation of the problem.

\section*{Acknowledgment}

This work was presented at 7th Marsel Grossmann Meeting \cite{27} due to the
support of International Science Foundation.

\section*{Appendix A}

\setcounter{equation}{0} \renewcommand{\theequation}{A.\arabic{equation}}

Here we will consider the expression (\ref{50}) for the function $F_{\alpha
=1/2}(\beta )$:

\begin{equation}
\label{A.1}F_{\alpha =1/2}(\beta )=-\frac 1{24\pi }+\frac{\pi \beta ^2}{12}%
+4\pi \beta ^2\stackrel{\infty }{\stackunder{n=0}{\sum }}\,\frac{n+1/2}{%
e^{4\pi ^2\beta (n+1/2)}-1}\,,
\end{equation}
and will obtain the other representation which is more convenient for an
analysis.

Let us rewrite the series in (A.1) in the other form:
\begin{equation}
\label{A.2}\stackrel{\infty }{\stackunder{n=0}{\sum }}\,\frac{n+1/2}{e^{4\pi
^2\beta (n+1/2)}-1}=\frac 12\stackunder{y\rightarrow 0}{\lim }\stackrel{%
\infty }{\,\stackunder{n=0}{\sum }}\,\frac{(2n+1)\cos (2n+1)y}{e^{2\pi
^2\beta (2n+1)}-1}\,.
\end{equation}
(We can do it because the functional series in the right-hand side of the
expression (A.2) is uniformly converged for any values of $y.$) Now the next
transformation can be executed:
\begin{equation}
\label{A.3}\stackrel{\infty }{\,\stackunder{n=0}{\sum }}\,\frac{(2n+1)\cos
(2n+1)y}{e^{2\pi ^2\beta (2n+1)}-1}=-\frac d{dy}\,\,\left[ \stackrel{\infty
}{\,\stackunder{n=0}{\sum }}\,\frac{q^{2n+1}\sin (2n+1)y}{1-q^{2n+1}}\right]
\,,
\end{equation}
where $q\equiv e^{-2\pi ^2\beta }$. Then we will use the formula (16.23.10)
from Ref.\cite{20}:
\begin{equation}
\label{A.4}\stackrel{\infty }{\,\stackunder{n=0}{\sum }}\,\frac{q^{2n+1}\sin
(2n+1)y}{1-q^{2n+1}}=\frac 14\csc y-\frac K{2\pi }\limfunc{ns}\left( \frac{2K%
}\pi y|m\right) \,.
\end{equation}
Here and below $\limfunc{ds}(z|m),\,\limfunc{cn}(z|m),\,\limfunc{ns}(z|m),\,
\limfunc{dn}(z|m),\,\limfunc{cn}(z|m),\,\limfunc{sn}(z|m)$ are the Jacobian
elliptic functions; $m$ is a parameter, $0\leq m\leq 1$; $K\equiv
K(m)=\int_0^{\pi /2}d\theta (1-m\sin {}^2\theta )^{-1/2}$ is the complete
elliptic integral \cite{24}. The parameters $m$ and $q$ are connected by the
relation
\begin{equation}
\label{A.5}q=\exp (-\pi K(1-m)/K(m))\,.
\end{equation}
Taking into account that $q=e^{-2\pi ^2\beta }$ we can obtain the connection
between parameters $m$ and $\beta $:
\begin{equation}
\label{A.6}2\pi \beta =K(1-m)/K(m)\,.
\end{equation}
Let us write out the differentiation formulae of the Jacobian functions \cite
{24}:
\begin{equation}
\label{A.7}
\begin{array}{c}
\limfunc{ds}{}^{\prime }(z|m)=-\limfunc{cs}(z|m)\limfunc{ns}(z|m)\,;\,\,\,\,
\limfunc{cs}{}^{\prime }(z|m)=-\limfunc{ns}(z|m)\limfunc{ds}(z|m)\,; \\
\limfunc{ns}{}^{\prime }(z|m)=-\limfunc{ds}(z|m)\limfunc{cs}(z|m)\,;
\end{array}
\end{equation}
take into account the relations \cite{24}:
\begin{equation}
\label{A.8}\limfunc{ds}(z|m)=\frac{\limfunc{dn}(z|m)}{\limfunc{sn}(z|m)}%
\,;\,\,\,\limfunc{cs}(z|m)=\frac{\limfunc{cn}(z|m)}{\limfunc{sn}(z|m)}\,;
\end{equation}
and allow for the expansion of elliptic functions in the degrees of $z$ \cite
{24}:%
$$
\limfunc{dn}(z|m)=1-m\frac{z^2}{2!}+O(z^4)\,;\,\,\,\limfunc{cn}(z|m)=1-\frac{%
z^2}{2!}+O(z^4)\,;
$$
\begin{equation}
\label{A.9}\limfunc{sn}(z|m)=z-(1+m)\frac{z^3}{3!}+O(z^5)\,.
\end{equation}
Now, substituting (A.2) into (A.1) and using consecutively the relations
(A.3-9) one can easy obtain the next representation for the function $%
F_{\alpha =1/2}(\beta )$:
\begin{equation}
\label{A.10}F_{\alpha =1/2}(\beta )=-\frac 1{24\pi }+\frac 1{3\pi }K^2\beta
^2(1+m)\,.
\end{equation}

In order to determine the region of change of the function $F_{\alpha
=1/2}(\beta )$ we take into account that the function $K(m)$ is
monotonically increasing from $\frac \pi 2$ to $+\infty $ if $m$ changes
from $0$ to $1$. Then, it is easy to see from (A.6) that the quantity $\beta
K(m)=K(1-m)/2\pi $ takes its values within an interval $(+\infty ,\frac 14)$%
. And finally, we can see from (A.10) that the function $F_{\alpha
=1/2}(\beta )$ is positive defined and its region of change is an interval
from $+\infty $ to 0.

\section*{Appendix B}

\setcounter{equation}{0} \renewcommand{\theequation}{B.\arabic{equation}}

Let us consider the expression (41) for the function $F_\alpha (\beta )$:%
$$
F_\alpha (\beta )=-\frac 1{24\pi }+\pi \beta ^2\left[ \frac 1{12}-\left(
\alpha -\frac 12\right) ^2\right] +
$$
\begin{equation}
\label{B.1}+2\pi \beta ^2\sum_{n=0}^\infty \frac{n+\alpha }{e^{4\pi ^2\beta
(n+\alpha )}-1}+2\pi \beta ^2\sum_{n=0}^\infty \frac{n+\overline{\alpha }}{%
e^{4\pi ^2\beta (n+\overline{\alpha })}-1}\,,
\end{equation}
where $\overline{\alpha }=1-\alpha $. The series in (B.1) are converged
fastly for big values of $\beta $ and slowly for small ones. With the aim to
make the numerical analysis of the function $F_\alpha (\beta )$ we will
obtain the other representation for the expression (B.1).

By using the simpliest form of the Euler-Maclaurin formula for a summation
\cite{25} we can rewrite the series in (B.1) as follows%
$$
\sum_{n=0}^\infty \frac{n+\alpha }{e^{4\pi ^2\beta (n+\alpha )}-1}%
=\int\limits_0^\infty dn\frac{n+\alpha }{e^{4\pi ^2\beta (n+\alpha )}-1}%
+\frac 12\frac \alpha {e^{4\pi ^2\beta \alpha }-1}+
$$
\begin{equation}
\label{B.2}+\int\limits_0^\infty dn\{n-[n]-1/2\}\frac d{dn}\frac{n+\alpha }{%
e^{4\pi ^2\beta (n+\alpha )}-1}\,.
\end{equation}
It is not difficult to transform the expression (B.2) into the next form:
\begin{equation}
\label{B.3}\frac{n+\alpha }{e^{4\pi ^2\beta (n+\alpha )}-1}=\frac 1{96\pi
^2\beta ^2}+\int\limits_0^\infty dx\{x-\alpha -[x-\alpha ]-1/2\}\frac
d{dx}\frac x{e^{4\pi ^2\beta x}-1}\,.
\end{equation}
So $F_\alpha (\beta )$ is%
$$
F_\alpha (\beta )=\pi \beta ^2\left[ \frac 1{12}-\left( \alpha -\frac
12\right) ^2\right] +
$$
\begin{equation}
\label{B.4}+2\pi \beta ^2\int\limits_0^\infty dx\left\{ (x-\alpha -[x-\alpha
]-1/2)+(x-\overline{\alpha }-[x-\overline{\alpha }]-1/2)\right\} \frac
d{dx}\frac x{e^{4\pi ^2\beta x}-1}\,.
\end{equation}
The periodic function $S(x)=x-[x]-1/2$ possesses the Fourier series
representation \cite{26}:
\begin{equation}
\label{B.5}S(x)=x-[x]-1/2=-\sum_{n=1}^\infty \frac{\sin 2\pi nx}{\pi n}\,.
\end{equation}
By using this representation we can transform the integrals in (B.4) as
follows%
$$
2\pi \beta ^2\int\limits_0^\infty dx\left\{ (x-\alpha -[x-\alpha ]-1/2)+(x-
\overline{\alpha }-[x-\overline{\alpha }]-1/2)\right\} \frac d{dx}\frac
x{e^{4\pi ^2\beta x}-1}=
$$
$$
=\frac \beta {2\pi }\int\limits_0^\infty dy\left\{ (z-\alpha -[z-\alpha
]-1/2)+(z-\overline{\alpha }-[z-\overline{\alpha }]-1/2)\right\} \frac
d{dy}\frac y{e^y-1}=
$$
\begin{equation}
\label{B.6}=-\frac \beta {2\pi }\sum_{n=1}^\infty \int\limits_0^\infty dy
\frac{\sin 2\pi n(z-\alpha )+\sin 2\pi n(z-\overline{\alpha })}{\pi n}\frac
d{dy}\frac y{e^y-1}\,,
\end{equation}
where $z=y/4\pi ^2\beta $. Substituting the expression (B.6) into (B.4) and
computing the integral we obtain the next expression for $F_\alpha (\beta )$:%
$$
F_\alpha (\beta )=\pi \beta ^2\left[ \frac 1{12}-\left( \alpha -\frac
12\right) ^2\right] +\frac{\beta ^2}\pi \sum_{n=1}^\infty \frac{\cos 2\pi
n\alpha }{n^2}-
$$
\begin{equation}
\label{B.7}-\frac 1{4\pi }\sum_{n=1}^\infty \frac{\cos 2\pi n\alpha }{\sinh
{}^2(n/2\beta )}\,.
\end{equation}
Using the formula (2.5.34) from Ref.\cite{20}:
\begin{equation}
\label{B.8}\sum_{n=1}^\infty \frac{\cos nx}{n^2}=\frac 14x^2-\frac{\pi x}2+
\frac{\pi ^2}6\,,
\end{equation}
we have finally
\begin{equation}
\label{B.9}F_\alpha (\beta )=-\frac 1{4\pi }\sum_{n=1}^\infty \frac{\cos
2\pi n\alpha }{\sinh {}^2(n/2\beta )}\,.
\end{equation}
The representation (B.9) for the function $F_\alpha (\beta )$ contains the
series which is fastly convered for small values of $\beta $.

Note also that in the case $\alpha =0$ (nontwisted scalar field)
\begin{equation}
\label{B.10}F_{\alpha =0}(\beta )=-\frac 1{4\pi }\sum_{n=1}^\infty \frac
1{\sinh {}^2(n/2\beta )}\,,
\end{equation}
and this expression coincides with one obtained in Ref.\cite{6}.

\newpage\

\end{document}